%% file: rasgf.tex
\documentclass[12pt]{article}
\usepackage{epsfig}
\begin{document}

\centerline{\bf\Large{How flat is the Universe?}}

\vspace*{1.00 cm}

\centerline{Matts Roos \& S. M. Harun-or-Rashid} 
\centerline{Department of Physics, Division of High Energy Physics,} 
\centerline{University of Helsinki, Finland.} 

\vspace*{1.00 cm}

\centerline {\bf Abstract}

In order to answer this question, we combine ten independent astrophysical
constraints in the space of the density parameters $\Omega_m$ of 
gravitating matter and $\Omega_{\Lambda}$ of vacuum energy. We find that 
$\Omega_m=0.31\pm 0.07$, $\Omega_{\Lambda}=0.63\pm 0.21$, and thus
$\Omega_m + \Omega_{\Lambda}=0.94\pm 0.22$. The total $\chi^2$ 
is 4.1 for 8 degrees of freedom, testifying that the various systematic
errors included are generous.  We also determine $\Omega_m$ in the 
exactly flat case. Five supplementary flat-case constraints can then
be included in our fit, with the result  $\Omega_m=1-\Omega_{\Lambda}=
0.337\pm0.031$. It follows that 
the age of the Universe is $t_0 = 13.5\pm 1.3$ (0.68/h) Gyr. \\


\section{Introduction}

If the dynamical parameters describing the cosmic expansion were known
to good precision, we would know whether the Universe is open or closed,
or whether its geo\-metry is in fact exactly flat as inflationary theory
wants it. To know the answer we need at least (i) the Hubble constant
$H_0$, usually given in the form $H_0 = 100h$ km s$^{-1}$ Mpc$^{-1}$,
(ii) the dimensionless density parameter $\Omega_m$ of gravitating
matter, comprising baryons, neutrinos and some yet unknown kinds of dark
matter, and (iii) the density parameter $\Omega_{\Lambda}$ of vacuum
energy, related to the cosmological constant $\Lambda$ by 
\begin{eqnarray}
\Omega_{\Lambda} = \Lambda/3H^2_0\ .
\label{f1}\end{eqnarray}         
A flat universe is defined by the condition
\begin{eqnarray}
\Omega_m + \Omega_{\Lambda}=1\ .
\label{f2}\end{eqnarray}

When $H_0, \Omega_m$ and $\Omega_{\Lambda}$ are known, the age of the
Universe, $t_0$, can be obtained from the Friedman-Lemaître model as
\begin{eqnarray}
t_0={{1}\over{H_0}}\int_0^1\hbox{d}x
\big{[}(1-\Omega_m-\Omega_{\Lambda})\nonumber \\ 
+\Omega_mx^{-1}+\Omega_{\Lambda}x^2
\big{]}^{-1/2}.  
\label{f3}\end{eqnarray}

In a previous publication (Roos \& Harun-or-Rashid 1998) we tried to
determine the preferred region in the $(\Omega_m,
\Omega_{\Lambda})$-plane by combining three independent observational
constraints and a value for $H_0$. Over the years more observational
constraints have become available, (some of them summarized in our 
unpublished preprint Roos \& Harun-or-Rashid 1999) so that we now 
can make use of 
fifteen independent constraints meeting our criteria. Since we combine 
the data in a least-squares fit, we can only make use of observations 
quoting a value and an error, but in addition many interesting limits 
also exist.

In Section 2 we describe the fifteen observational constraints entering our
least-squares fit.  Ten constraints are valid in the space of $\Omega_m$ and
$\Omega_{\Lambda}$ ; the remaining five are only valid on the flat line 
Eq.  (\ref{f2}) and will be included only when we fix the fit to that line.  
In Section 3 we describe the results of our fit 
in the $(\Omega_m, \Omega_{\Lambda})$-plane as well as
along the flat line Eq.  (\ref{f2}). We then also use Eq.  (\ref{f3})
to determine $t_0$. 

\section{Observational Constraints}

For the Hubble constant we use the value 
$h=0.68\pm 0.05$ from the analysis of Nevalainen \& Roos (1998) in
which the Cepheid period-luminosity relation is corrected for metallicity
dependence. This agrees well with more recent precise determinations. For
instance Mould \& al (1999) find, when similarly corrected, h = 0.67 
(our evaluation).

\subsection{Cosmic Microwave Background Radiation}
The observations of anisotropies in the CMBR are commonly presented as
plots of the multipole moments $C_{\ell}$ against the multipole $\ell$,
or equivalently, against the FWHM value of the angular anisotropy
signal. In general, the theoretical models for the power
spectrum may depend on up to 9 parameters.  Lineweaver (1998)
and Tegmark (1998) have combined the data from MAP and PLANCK into a
confidence region in the marginal subspace
of the $(\Omega_m, \Omega_{\Lambda})$-plane.  
An independent analysis, combining the angular power spectrum of the 
BOOMERANG experiment (Melchiorri et al. 1999) with that of COBE, furnishes
us a second constraint.

\subsection{Gas fraction in X-ray clusters}

Matter in an idealized, spherically symmetric cluster is taken to be
made up of a nearly hydrostatic inner body surrounded by an outer envelope, 
infalling with the cosmic mix of the components. The
baryonic component of the mass in galaxy clusters is dominated by gas
which can be observed by its X-ray emission. Thus by measuring the gas
fraction, one expects to obtain fairly unbiased
information on the ratio of $\Omega_m$ to the cosmic baryonic density
parameter $\Omega_b$. Using a very large sample of clusters, Evrard (1997)
has obtained a 'realistic' value of 
\begin{eqnarray}
{\Omega_m\over \Omega_b} h^{-4/3} \approx (11.8\pm 0.7)\ .
\label{f4}\end{eqnarray}

Taking $\Omega_b=0.024\pm 0.006 h^{-2}$ from the low primordial
deuterium abundance (Tytler, Fan \& Burles 1996), one obtains 
\begin{eqnarray}
\Omega_m=0.36\pm 0.09\ , 
\label{f5}\end{eqnarray}
which we use as our constraint.

This constraint are restricted to a flat Universe and we only use
them together with the assumption of flat cosmology in section 3.

\subsection{Cluster mass function and the Ly$\alpha$ forest}

In theories of structure formation based on gravitational instability
and Gaussian initial fluctuations, massive galaxy clusters can form
either by the collapse of large volumes in a low density universe, or by
the collapse of smaller volumes in a high density universe. This is
expressed by the cluster mass function which constrains a combination
of  $\Omega_m$ and the amplitude $\sigma$ of mass fluctuations
(normalized inside some volume). The amplitude $\sigma$ is given by an
integral over the mass power spectrum. From an analysis by Weinberg \& al.
(1998) combining the 
cluster mass function constraint with the linear mass power spectrum  
determined from Ly$\alpha$ data, one obtains the relation
\begin{eqnarray}
\Omega_m +  0. 18 \Omega_{\Lambda}=0.46\pm 0.08
\label{f6}\end{eqnarray}
which we use as one constraint.

\subsection{X-ray cluster evolution} 

Clusters of galaxies are the largest known gravitationally bound
structures in the Universe. Since they are thought to be formed by
contraction from density fluctuations in an initially fairly homogeneous
Universe, their distribution in redshift and their density spectrum as
seen in their X-ray emission gives precious information about their
formation and evolution with time.  Thus by combining the evolution in
abundance of X-ray clusters with their luminosity-temperature
correlation, one obtains a powerful test of the mean density of the
Universe.

The results of Bahcall, Fan and Cen (1997) can be summarized in the relation 
\begin{eqnarray}
\Omega_m =  0.195\pm 0.11 + 0.071 \Omega_{\Lambda}
\label{f7}\end{eqnarray}
which we use as one constraint.

The results of Eke \& al. (1998) can be summarized in the relation 
\begin{eqnarray}
\Omega_m =  0.44\pm 0.20 - 0.077 \Omega_{\Lambda}
\label{f8}\end{eqnarray}
which we use as one constraint.

Donahue \& Voit (1999) constrain $\Omega_m$ through a maximum likelihood
analysis of teperatures and redshifts of the high redshift clusters from 
the Extended Medium Redshift Survey, as well as from a low redshift sample
(Markevitch 1998), finding 
\begin{eqnarray}
\Omega_m=0.27\pm 0.10\ , 
\label{f5a}\end{eqnarray}
This constraint are restricted to a flat Universe and we only use
them together with the assumption of flat cosmology in section 3.

\subsection{Gravitational lensing}

The number of multiply imaged QSOs found in lens surveys is sensitive
to  $\Omega_{\Lambda}$. Models of gravitational lensing must, however,
explain not only the observed probability of lensing, but also the
relative probability of showing a specific image separation. The image
separation increases with increasing $\sigma^*$, the characteristic
velocity dispersion. Thus the results can be expressed as likelihood
contour plots in the two-dimensional parameter space of $\sigma^*$ and
$\Omega_m=1-\Omega_{\Lambda}$ (Chiba \& Yoshii 1999). 

We integrate out $\sigma^*$, and we thus obtain the constraint 
\begin{eqnarray}
\Omega_{\Lambda}=0.70\pm 0.16\
\label{f9}\end{eqnarray}

Im, Griffiths \& Ratnatunga (1997) use seven field elliptical galaxies
to determine
\begin{eqnarray}
\Omega_{\Lambda}=0.64\pm 0.15\
\label{f9a}\end{eqnarray}
in the flat model.

We use these two constraints together with the assumption of a flat 
cosmology in section 3.

\subsection{Classical double radio sources}

There are two independent measures of the average size of a radio
source, where size implies the separation of two hot spots: the average
size of similar sources at the same redshift, and the product of the
average rate of growth of the source and the total time for which the
highly collimated outflows of that source are powered by the AGN. This
outflow leads to the large scale radio emission. The two measures depend
on the angular size distance to the source in different ways, so
equating them allows a determination of the coordinate distance to the
source which, in turn, can be used to determine pairs of $\Omega_m,
\Omega_{\Lambda}$-values.  

By using 14 classical double radio galaxies, Daly, Guerra \& Wan Lin 
(1998) determine an approximately
elliptical 68\% confidence region in the ($\Omega_m,
\Omega_{\Lambda}$)-plane centered at (0.05, 0.32). In our fit we use this
constraint.

\subsection{Supernov\ae\ of type Ia}

Type Ia supernovae can be calibrated as standard candles, and have
enormous luminosities. 
These two features make them a near-ideal tool for studying the
luminosity-redshift relationship at cosmological distances.  

The factor relating brightness to redshift is a function of $\Omega_m$ 
and $\Omega_\Lambda$.  
The High-z Supernova Search Team Riess et al. (1998) have used 10 SNe
Ia in the redshift range 0.16 -- 0.62 to
place constraints on the dynamical parameters and $t_0$.  In the
($\Omega_m, \Omega_{\Lambda}$)-plane their 68\% confidence region is an
ellipse centered at (0.20, 0.65) which we use as one constraint.

The Supernova Cosmology Project (Perlmutter et al. 1998) has published 
an analysis based
on 42 supernov\ae\ in the high-redshift range 0.18 -- 0.83. In the 
($\Omega_m, \Omega_{\Lambda}$)-plane the 68\% confidence range is
an ellipse centered at (0.75, 1.36) which we use as one constraint.

\subsection{Power-spectrum of extragalactic objects}

Matter in every direction appear to be distributed in high-density
peaks sepatated by voids. The average separation distance is 
$\sim130h^{-1}$Mpc, which translates into a peak in the power spectrum of 
mass fluctuations. This provides a co-moving scale for measuring 
cosmological curvature. Broadhurst \& Jaffe (1999) used a set of 
Lyman galaxies at $z\sim 3$ finding a constraint of the form
\begin{eqnarray}
\Omega_m =  0.20\pm 0.10 + 0.34 \Omega_{\Lambda}
\label{f10}\end{eqnarray}

Roukema \& Mamon (1999) have carried out a similar analysis of quasars, 
finding
\begin{eqnarray}
\Omega_m =  0.24\pm 0.15 + (0.10\pm 0.08)\Omega_{\Lambda}
\label{f10a}\end{eqnarray}

We use these two results as constraints. \\

\subsection{Galaxy peculiar velocities}

The large-scale peculiar velocities of galaxies correspond via gravity to
mass density fluctuations about the mean, and depend also on the mean
density itself. Two catalogs of galaxies have been analyzed for these
velocities in order to provide information on $\Omega_m$: the Mark III
catalog (Willick \& al. 1997) of about 3000 galaxies within a distance of
$\sim 70 h^{-1}$ Mpc, and the SFI catalog (Borgani \& al. 1999) of about
1300 spiral galaxies in a similar volume. Combining the results in these
catalogs, Zehavi \& Dekel (1999) quote the constraint
\begin{eqnarray}
\Omega_m h_{65}^{1.3}n^2\simeq 0.58\pm 0.12\ ,
\label{f17}\end{eqnarray}
in the case of flat cosmology,
where the error corresponds to a 90\% confidence range. Taking the index
$n$ of the mass-density fluctuation power spectrum to be $n=1.0\pm 0.1$
(Bond \& Jaffe 1998), one obtains the constraint
\begin{eqnarray}
\Omega_m= 0.55\pm 0.14\ ,
\label{f18}\end{eqnarray}
where the error corresponds to a 68\% confidence range.
This constraint we only use together with the assumption of a flat 
cosmology in section 3.

\section{Results and Discussion}

\subsection{Fits}

We perform a least-squares fit to the above ten constraints in the space 
of the two free parameters $\Omega_m, \Omega_{\Lambda}$.  The Hubble 
constant is not treated as a free parameter, but is fixed to the
value Nevalainen \& Roos (1998).  

Paying rigorous attention to statistical detail, we use the standard 
minimization program MINUIT (James \& Roos 1975). The best fit value is 
then found to be 
\begin{eqnarray} \Omega_m=0.31\pm
0.07,\ \ \Omega_{\Lambda}=0.63\pm 0.21, \nonumber \\ 
\chi^2=4.1\ . 
\label{f11}\end{eqnarray} 
In Fig.1 we plot the shape of the $1\sigma$ and $2\sigma$ contours.  

The above values can be added to yield
\begin{eqnarray}
\Omega_m + \Omega_{\Lambda}=0.94\pm 0.22\ .
\label{f12}\end{eqnarray}

From this we conclude that (i) the data require a flat cosmology, (ii)
the Einstein-de Sitter model is very convincingly ruled out, and (iii)
also any low-density model with $\Omega_{\Lambda}=0$ is ruled out. 
 
If we assume exact flatness and refit the previous ten constraints as well 
as the five one-dimensional constraints (5),(6),(10),(11),(15), the result is

\begin{eqnarray} 
\Omega_m=0.337\pm 0.031 ,\ \ \Omega_{\Lambda}=0.663\pm 0.031,\nonumber \\
 \chi^2=7.3\ .  
\label{f13}\end{eqnarray}

The value of $\Omega_{\Lambda}$ in the two-dimensional fit, Eq. (\ref{f11}),
is determined mainly by the constraint from the Supernovae Cosmology
Project (Perlmutter et al. 1998). However, on the flat line several other
constraints contribute much more strongly, so that the result in 
Eq. (\ref{f13}) is very precise, even indepently of the supernovae
constraint.

Let us now substitute the above parameter values into
Eq. (\ref{f3}). Adding a 7.4\% $H_0$ error quadratically to the 
density parameter errors
in Eq. (\ref{f11}), propagated through the integral (\ref{f3}), we find
as a value for the age of the Universe
\begin{eqnarray}
t_0 = 13.5 \pm 1.3\ (0.68/h)\  \rm{Gyr}\ .
\label{f14}\end{eqnarray}
For an exactly flat Universe, only the error changes slightly to 1.1.

\subsection{Systematic errors}

With results as precise as those for the flat model, a question arising
is, what about neglected systematic errors? 

The constraints we use are indeed pulling the results in every
direction. CMBR is orthogonal to the supernova constraints, the
gravitational lensing constraint is exactly orthogonal to
the gas fraction in X-ray clusters and the remaining
constraints represent bands of several different directions.
Thus we think that it is justified to consider the
systematic errors as random and the total effect of possibly neglected
systematic errors to be mutual cancellation.

Moreover, the two-dimensional fit,
Eq. (\ref{f11}), $\chi^2$ is 4.1 for 8 degrees of freedom, and in the
one-dimensional fit, Eq. (\ref{f13}), $\chi^2$ is 7.3 for 14 degrees of
freedom, much too low for statistically
distributed data. Thus we can conclude that the various errors quoted
for our fifteen constraints are not statistical: they have been blown up
unreasonably by the systematic errors added, and there is no motivation 
for blowing them up further by adding arbitrary systematic errors.

\subsection{Comparison with other data}

There are some categories of data which we have not used, but to which
it is nevertheless interesting to compare our results.

Totani, Yoshii \& Sato (1997) have tested cosmological models for the
evolution of galaxies and star creation against the evolution of galaxy
luminosity densities. They have found $\Omega_{\Lambda}>0.53$
at 95\% confidence in a flat universe.

Falco, Kochanek \& Munoz (1998) have determined the redshift
distribution of 124 radio sources and used it to derive a limit on
$\Omega_m$ from the statistics of six gravitational lenses. In a flat
universe their best fit yields $\Omega_m>0.26$ at 95.5\% confidence.

Since we also determine a value for the age of the Universe, it is of
interest to look at other recent $t_0$ determinations. By using several 
techniques Chaboyer (1998) determine the
age of the oldest globular clusters (GC) to $t_{GC}=11.5\pm 1.3$ Gyr. 
Another estimate of the age of the oldest globular clusters is due to
Jimenez (1998). He quotes the 99\% confidence range $t_{GC}=13.25\pm
2.75$ Gyr, which translates into the 68\% confidence range
$t_{GC}=13.3\pm 1.1$ Gyr. Taking a mid-value and adding 0.9 Gyr systematic
error to account for the spread of values, we have 

\begin{eqnarray}
t_{GC} = 12.4\pm 1.3\pm 0.9\ \rm{Gyr}\ .
\label{f19}\end{eqnarray}

In order to obtain $t_0$ from the GC value, however, one must add the
time it took for the metal-poor stars to form. Estimates are poor due to
the lack of a good theory, so Chaboyer (1998) recommends adding 0.1 to 2
Gyr. We presume that this is a 90\% CL estimate, $+1\pm 1$ Gyr. Thus one
arrives at the 68\% confidence range 
\begin{eqnarray}
t_0 = 13.4\pm 1.3\pm 0.9\pm 0.6\ \rm{Gyr} = 13.4\pm 1.7\ \rm{Gyr}\ .
\label{f21}\end{eqnarray}

Jimenez (1998) also finds an age of $t_0 = 13\pm 2$ Gyr for the galaxy 
53W069 at $z=1.43$. This value as well as Eq. (\ref{f21}) are in excellent 
agreement with our value in Eq. (\ref{f14})

\section* {Acknowledgements} 

The authors wish to thank J. Nevalainen, Helsinki, for useful comments. 
S.M.H. is indebted to the Magnus Ehrnrooth Foundation for support.


\begin{figure}
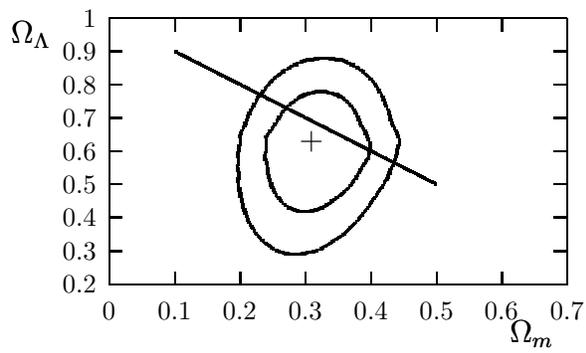

\vspace{0pc}
\input sig12.tex
\vspace{0.5cm}
\caption{ The 1$\sigma$ and 2$\sigma$ statistical 
confidence regions in the ($\Omega_m,\Omega_{\Lambda}$)-plane 
are shown. The '+' marks the best fit: ($\Omega_m$,$\Omega_{\Lambda}$) 
= (0.31,0.63). The diagonal line corresponds to a flat cosmology.}
\label{appenfig}
\end{figure}


\label{lastpage}

\end{document}

%% file: sig12.tex
\setlength{\unitlength}{0.240900pt}
\ifx\plotpoint\undefined\newsavebox{\plotpoint}\fi
\begin{picture}(900,540)(0,0)
\font\gnuplot=cmr10 at 10pt
\gnuplot
\sbox{\plotpoint}{\rule[-0.200pt]{0.400pt}{0.400pt}}%
\put(120.0,82.0){\rule[-0.200pt]{4.818pt}{0.400pt}}
\put(100,82){\makebox(0,0)[r]{0.2}}
\put(819.0,82.0){\rule[-0.200pt]{4.818pt}{0.400pt}}
\put(120.0,134.0){\rule[-0.200pt]{4.818pt}{0.400pt}}
\put(100,134){\makebox(0,0)[r]{0.3}}
\put(819.0,134.0){\rule[-0.200pt]{4.818pt}{0.400pt}}
\put(120.0,187.0){\rule[-0.200pt]{4.818pt}{0.400pt}}
\put(100,187){\makebox(0,0)[r]{0.4}}
\put(819.0,187.0){\rule[-0.200pt]{4.818pt}{0.400pt}}
\put(120.0,239.0){\rule[-0.200pt]{4.818pt}{0.400pt}}
\put(100,239){\makebox(0,0)[r]{0.5}}
\put(819.0,239.0){\rule[-0.200pt]{4.818pt}{0.400pt}}
\put(120.0,291.0){\rule[-0.200pt]{4.818pt}{0.400pt}}
\put(100,291){\makebox(0,0)[r]{0.6}}
\put(819.0,291.0){\rule[-0.200pt]{4.818pt}{0.400pt}}
\put(120.0,343.0){\rule[-0.200pt]{4.818pt}{0.400pt}}
\put(100,343){\makebox(0,0)[r]{0.7}}
\put(819.0,343.0){\rule[-0.200pt]{4.818pt}{0.400pt}}
\put(120.0,395.0){\rule[-0.200pt]{4.818pt}{0.400pt}}
\put(100,395){\makebox(0,0)[r]{0.8}}
\put(819.0,395.0){\rule[-0.200pt]{4.818pt}{0.400pt}}
\put(120.0,448.0){\rule[-0.200pt]{4.818pt}{0.400pt}}
\put(100,448){\makebox(0,0)[r]{0.9}}
\put(819.0,448.0){\rule[-0.200pt]{4.818pt}{0.400pt}}
\put(120.0,500.0){\rule[-0.200pt]{4.818pt}{0.400pt}}
\put(100,500){\makebox(0,0)[r]{1}}
\put(819.0,500.0){\rule[-0.200pt]{4.818pt}{0.400pt}}
\put(120.0,82.0){\rule[-0.200pt]{0.400pt}{4.818pt}}
\put(120,41){\makebox(0,0){0}}
\put(120.0,480.0){\rule[-0.200pt]{0.400pt}{4.818pt}}
\put(223.0,82.0){\rule[-0.200pt]{0.400pt}{4.818pt}}
\put(223,41){\makebox(0,0){0.1}}
\put(223.0,480.0){\rule[-0.200pt]{0.400pt}{4.818pt}}
\put(325.0,82.0){\rule[-0.200pt]{0.400pt}{4.818pt}}
\put(325,41){\makebox(0,0){0.2}}
\put(325.0,480.0){\rule[-0.200pt]{0.400pt}{4.818pt}}
\put(428.0,82.0){\rule[-0.200pt]{0.400pt}{4.818pt}}
\put(428,41){\makebox(0,0){0.3}}
\put(428.0,480.0){\rule[-0.200pt]{0.400pt}{4.818pt}}
\put(531.0,82.0){\rule[-0.200pt]{0.400pt}{4.818pt}}
\put(531,41){\makebox(0,0){0.4}}
\put(531.0,480.0){\rule[-0.200pt]{0.400pt}{4.818pt}}
\put(634.0,82.0){\rule[-0.200pt]{0.400pt}{4.818pt}}
\put(634,41){\makebox(0,0){0.5}}
\put(634.0,480.0){\rule[-0.200pt]{0.400pt}{4.818pt}}
\put(736.0,82.0){\rule[-0.200pt]{0.400pt}{4.818pt}}
\put(736,41){\makebox(0,0){0.6}}
\put(736.0,480.0){\rule[-0.200pt]{0.400pt}{4.818pt}}
\put(839.0,82.0){\rule[-0.200pt]{0.400pt}{4.818pt}}
\put(839,41){\makebox(0,0){0.7}}
\put(839.0,480.0){\rule[-0.200pt]{0.400pt}{4.818pt}}
\put(120.0,82.0){\rule[-0.200pt]{173.207pt}{0.400pt}}
\put(839.0,82.0){\rule[-0.200pt]{0.400pt}{100.696pt}}
\put(120.0,500.0){\rule[-0.200pt]{173.207pt}{0.400pt}}
\put(818,4){\makebox(0,0)[r]{$\Omega_m$}}
\put(28,474){\makebox(0,0)[r]{$\Omega_\Lambda$}}
\put(818,4){\makebox(0,0)[r]{$\Omega_m$}}
\put(28,474){\makebox(0,0)[r]{$\Omega_\Lambda$}}
\put(120.0,82.0){\rule[-0.200pt]{0.400pt}{100.696pt}}
\sbox{\plotpoint}{\rule[-0.400pt]{0.800pt}{0.800pt}}%
\put(432,196){\usebox{\plotpoint}}
\put(432,194.84){\rule{0.964pt}{0.800pt}}
\multiput(432.00,194.34)(2.000,1.000){2}{\rule{0.482pt}{0.800pt}}
\put(436,196.84){\rule{2.891pt}{0.800pt}}
\multiput(436.00,195.34)(6.000,3.000){2}{\rule{1.445pt}{0.800pt}}
\put(448,199.34){\rule{1.445pt}{0.800pt}}
\multiput(448.00,198.34)(3.000,2.000){2}{\rule{0.723pt}{0.800pt}}
\multiput(454.00,203.38)(1.096,0.560){3}{\rule{1.640pt}{0.135pt}}
\multiput(454.00,200.34)(5.596,5.000){2}{\rule{0.820pt}{0.800pt}}
\multiput(463.00,208.39)(0.574,0.536){5}{\rule{1.133pt}{0.129pt}}
\multiput(463.00,205.34)(4.648,6.000){2}{\rule{0.567pt}{0.800pt}}
\multiput(470.00,214.38)(0.928,0.560){3}{\rule{1.480pt}{0.135pt}}
\multiput(470.00,211.34)(4.928,5.000){2}{\rule{0.740pt}{0.800pt}}
\multiput(478.00,219.40)(0.599,0.514){13}{\rule{1.160pt}{0.124pt}}
\multiput(478.00,216.34)(9.592,10.000){2}{\rule{0.580pt}{0.800pt}}
\multiput(490.00,229.40)(0.481,0.520){9}{\rule{1.000pt}{0.125pt}}
\multiput(490.00,226.34)(5.924,8.000){2}{\rule{0.500pt}{0.800pt}}
\multiput(499.40,236.00)(0.514,0.654){13}{\rule{0.124pt}{1.240pt}}
\multiput(496.34,236.00)(10.000,10.426){2}{\rule{0.800pt}{0.620pt}}
\multiput(509.39,249.00)(0.536,1.020){5}{\rule{0.129pt}{1.667pt}}
\multiput(506.34,249.00)(6.000,7.541){2}{\rule{0.800pt}{0.833pt}}
\multiput(515.39,260.00)(0.536,0.909){5}{\rule{0.129pt}{1.533pt}}
\multiput(512.34,260.00)(6.000,6.817){2}{\rule{0.800pt}{0.767pt}}
\multiput(521.40,270.00)(0.526,0.825){7}{\rule{0.127pt}{1.457pt}}
\multiput(518.34,270.00)(7.000,7.976){2}{\rule{0.800pt}{0.729pt}}
\put(526.34,281){\rule{0.800pt}{2.409pt}}
\multiput(525.34,281.00)(2.000,5.000){2}{\rule{0.800pt}{1.204pt}}
\put(528.34,291){\rule{0.800pt}{2.409pt}}
\multiput(527.34,291.00)(2.000,5.000){2}{\rule{0.800pt}{1.204pt}}
\multiput(529.06,301.00)(-0.560,1.432){3}{\rule{0.135pt}{1.960pt}}
\multiput(529.34,301.00)(-5.000,6.932){2}{\rule{0.800pt}{0.980pt}}
\put(522.34,312){\rule{0.800pt}{2.200pt}}
\multiput(524.34,312.00)(-4.000,5.434){2}{\rule{0.800pt}{1.100pt}}
\multiput(520.08,322.00)(-0.526,0.825){7}{\rule{0.127pt}{1.457pt}}
\multiput(520.34,322.00)(-7.000,7.976){2}{\rule{0.800pt}{0.729pt}}
\put(511.34,333){\rule{0.800pt}{2.200pt}}
\multiput(513.34,333.00)(-4.000,5.434){2}{\rule{0.800pt}{1.100pt}}
\multiput(509.06,343.00)(-0.560,1.432){3}{\rule{0.135pt}{1.960pt}}
\multiput(509.34,343.00)(-5.000,6.932){2}{\rule{0.800pt}{0.980pt}}
\multiput(504.08,354.00)(-0.526,0.738){7}{\rule{0.127pt}{1.343pt}}
\multiput(504.34,354.00)(-7.000,7.213){2}{\rule{0.800pt}{0.671pt}}
\multiput(494.85,365.38)(-0.424,0.560){3}{\rule{1.000pt}{0.135pt}}
\multiput(496.92,362.34)(-2.924,5.000){2}{\rule{0.500pt}{0.800pt}}
\multiput(489.30,370.39)(-0.574,0.536){5}{\rule{1.133pt}{0.129pt}}
\multiput(491.65,367.34)(-4.648,6.000){2}{\rule{0.567pt}{0.800pt}}
\put(483,374.34){\rule{0.964pt}{0.800pt}}
\multiput(485.00,373.34)(-2.000,2.000){2}{\rule{0.482pt}{0.800pt}}
\put(477,376.84){\rule{1.445pt}{0.800pt}}
\multiput(480.00,375.34)(-3.000,3.000){2}{\rule{0.723pt}{0.800pt}}
\put(470,379.34){\rule{1.686pt}{0.800pt}}
\multiput(473.50,378.34)(-3.500,2.000){2}{\rule{0.843pt}{0.800pt}}
\put(466,380.84){\rule{0.964pt}{0.800pt}}
\multiput(468.00,380.34)(-2.000,1.000){2}{\rule{0.482pt}{0.800pt}}
\put(458,382.34){\rule{1.927pt}{0.800pt}}
\multiput(462.00,381.34)(-4.000,2.000){2}{\rule{0.964pt}{0.800pt}}
\put(439,382.34){\rule{2.168pt}{0.800pt}}
\multiput(443.50,383.34)(-4.500,-2.000){2}{\rule{1.084pt}{0.800pt}}
\put(435,380.84){\rule{0.964pt}{0.800pt}}
\multiput(437.00,381.34)(-2.000,-1.000){2}{\rule{0.482pt}{0.800pt}}
\put(428,379.34){\rule{1.686pt}{0.800pt}}
\multiput(431.50,380.34)(-3.500,-2.000){2}{\rule{0.843pt}{0.800pt}}
\put(421,376.84){\rule{1.686pt}{0.800pt}}
\multiput(424.50,378.34)(-3.500,-3.000){2}{\rule{0.843pt}{0.800pt}}
\put(417,374.34){\rule{0.964pt}{0.800pt}}
\multiput(419.00,375.34)(-2.000,-2.000){2}{\rule{0.482pt}{0.800pt}}
\multiput(412.30,373.07)(-0.574,-0.536){5}{\rule{1.133pt}{0.129pt}}
\multiput(414.65,373.34)(-4.648,-6.000){2}{\rule{0.567pt}{0.800pt}}
\multiput(403.86,367.06)(-0.928,-0.560){3}{\rule{1.480pt}{0.135pt}}
\multiput(406.93,367.34)(-4.928,-5.000){2}{\rule{0.740pt}{0.800pt}}
\multiput(400.08,359.48)(-0.516,-0.548){11}{\rule{0.124pt}{1.089pt}}
\multiput(400.34,361.74)(-9.000,-7.740){2}{\rule{0.800pt}{0.544pt}}
\multiput(391.08,347.95)(-0.526,-0.825){7}{\rule{0.127pt}{1.457pt}}
\multiput(391.34,350.98)(-7.000,-7.976){2}{\rule{0.800pt}{0.729pt}}
\multiput(384.07,336.63)(-0.536,-0.909){5}{\rule{0.129pt}{1.533pt}}
\multiput(384.34,339.82)(-6.000,-6.817){2}{\rule{0.800pt}{0.767pt}}
\multiput(378.08,327.60)(-0.520,-0.700){9}{\rule{0.125pt}{1.300pt}}
\multiput(378.34,330.30)(-8.000,-8.302){2}{\rule{0.800pt}{0.650pt}}
\multiput(370.08,316.43)(-0.526,-0.738){7}{\rule{0.127pt}{1.343pt}}
\multiput(370.34,319.21)(-7.000,-7.213){2}{\rule{0.800pt}{0.671pt}}
\put(448.0,385.0){\rule[-0.400pt]{2.409pt}{0.800pt}}
\put(362.84,281){\rule{0.800pt}{2.409pt}}
\multiput(363.34,286.00)(-1.000,-5.000){2}{\rule{0.800pt}{1.204pt}}
\put(364.34,270){\rule{0.800pt}{2.400pt}}
\multiput(362.34,276.02)(4.000,-6.019){2}{\rule{0.800pt}{1.200pt}}
\put(366.84,260){\rule{0.800pt}{2.409pt}}
\multiput(366.34,265.00)(1.000,-5.000){2}{\rule{0.800pt}{1.204pt}}
\put(368.34,249){\rule{0.800pt}{2.650pt}}
\multiput(367.34,254.50)(2.000,-5.500){2}{\rule{0.800pt}{1.325pt}}
\put(371.34,236){\rule{0.800pt}{2.800pt}}
\multiput(369.34,243.19)(4.000,-7.188){2}{\rule{0.800pt}{1.400pt}}
\put(375.34,228){\rule{0.800pt}{1.800pt}}
\multiput(373.34,232.26)(4.000,-4.264){2}{\rule{0.800pt}{0.900pt}}
\multiput(380.40,222.43)(0.526,-0.738){7}{\rule{0.127pt}{1.343pt}}
\multiput(377.34,225.21)(7.000,-7.213){2}{\rule{0.800pt}{0.671pt}}
\put(386.34,213){\rule{0.800pt}{1.200pt}}
\multiput(384.34,215.51)(4.000,-2.509){2}{\rule{0.800pt}{0.600pt}}
\multiput(391.38,208.18)(0.560,-0.592){3}{\rule{0.135pt}{1.160pt}}
\multiput(388.34,210.59)(5.000,-3.592){2}{\rule{0.800pt}{0.580pt}}
\multiput(395.00,205.06)(0.760,-0.560){3}{\rule{1.320pt}{0.135pt}}
\multiput(395.00,205.34)(4.260,-5.000){2}{\rule{0.660pt}{0.800pt}}
\put(402,199.34){\rule{1.686pt}{0.800pt}}
\multiput(402.00,200.34)(3.500,-2.000){2}{\rule{0.843pt}{0.800pt}}
\put(409,196.84){\rule{2.168pt}{0.800pt}}
\multiput(409.00,198.34)(4.500,-3.000){2}{\rule{1.084pt}{0.800pt}}
\put(418,194.84){\rule{0.964pt}{0.800pt}}
\multiput(418.00,195.34)(2.000,-1.000){2}{\rule{0.482pt}{0.800pt}}
\put(365.0,291.0){\rule[-0.400pt]{0.800pt}{5.059pt}}
\put(422.0,196.0){\rule[-0.400pt]{2.409pt}{0.800pt}}
\put(418,129){\usebox{\plotpoint}}
\put(418,127.84){\rule{1.927pt}{0.800pt}}
\multiput(418.00,127.34)(4.000,1.000){2}{\rule{0.964pt}{0.800pt}}
\put(426,129.34){\rule{1.686pt}{0.800pt}}
\multiput(426.00,128.34)(3.500,2.000){2}{\rule{0.843pt}{0.800pt}}
\put(433,131.34){\rule{2.409pt}{0.800pt}}
\multiput(433.00,130.34)(5.000,2.000){2}{\rule{1.204pt}{0.800pt}}
\multiput(443.00,135.38)(1.600,0.560){3}{\rule{2.120pt}{0.135pt}}
\multiput(443.00,132.34)(7.600,5.000){2}{\rule{1.060pt}{0.800pt}}
\multiput(455.00,140.40)(0.788,0.512){15}{\rule{1.436pt}{0.123pt}}
\multiput(455.00,137.34)(14.019,11.000){2}{\rule{0.718pt}{0.800pt}}
\multiput(472.00,151.40)(0.788,0.512){15}{\rule{1.436pt}{0.123pt}}
\multiput(472.00,148.34)(14.019,11.000){2}{\rule{0.718pt}{0.800pt}}
\multiput(489.00,162.40)(0.487,0.514){13}{\rule{1.000pt}{0.124pt}}
\multiput(489.00,159.34)(7.924,10.000){2}{\rule{0.500pt}{0.800pt}}
\multiput(499.00,172.40)(0.487,0.514){13}{\rule{1.000pt}{0.124pt}}
\multiput(499.00,169.34)(7.924,10.000){2}{\rule{0.500pt}{0.800pt}}
\multiput(510.40,181.00)(0.516,0.611){11}{\rule{0.124pt}{1.178pt}}
\multiput(507.34,181.00)(9.000,8.555){2}{\rule{0.800pt}{0.589pt}}
\multiput(519.40,192.00)(0.520,0.627){9}{\rule{0.125pt}{1.200pt}}
\multiput(516.34,192.00)(8.000,7.509){2}{\rule{0.800pt}{0.600pt}}
\multiput(527.40,202.00)(0.512,0.739){15}{\rule{0.123pt}{1.364pt}}
\multiput(524.34,202.00)(11.000,13.170){2}{\rule{0.800pt}{0.682pt}}
\multiput(538.41,218.00)(0.511,0.762){17}{\rule{0.123pt}{1.400pt}}
\multiput(535.34,218.00)(12.000,15.094){2}{\rule{0.800pt}{0.700pt}}
\multiput(550.40,236.00)(0.520,0.847){9}{\rule{0.125pt}{1.500pt}}
\multiput(547.34,236.00)(8.000,9.887){2}{\rule{0.800pt}{0.750pt}}
\put(557.34,249){\rule{0.800pt}{2.400pt}}
\multiput(555.34,249.00)(4.000,6.019){2}{\rule{0.800pt}{1.200pt}}
\multiput(562.38,260.00)(0.560,1.264){3}{\rule{0.135pt}{1.800pt}}
\multiput(559.34,260.00)(5.000,6.264){2}{\rule{0.800pt}{0.900pt}}
\put(566.34,270){\rule{0.800pt}{2.400pt}}
\multiput(564.34,270.00)(4.000,6.019){2}{\rule{0.800pt}{1.200pt}}
\put(569.34,281){\rule{0.800pt}{2.409pt}}
\multiput(568.34,281.00)(2.000,5.000){2}{\rule{0.800pt}{1.204pt}}
\put(571.34,291){\rule{0.800pt}{2.409pt}}
\multiput(570.34,291.00)(2.000,5.000){2}{\rule{0.800pt}{1.204pt}}
\put(570.84,312){\rule{0.800pt}{2.409pt}}
\multiput(572.34,312.00)(-3.000,5.000){2}{\rule{0.800pt}{1.204pt}}
\multiput(569.07,322.00)(-0.536,1.020){5}{\rule{0.129pt}{1.667pt}}
\multiput(569.34,322.00)(-6.000,7.541){2}{\rule{0.800pt}{0.833pt}}
\put(561.34,333){\rule{0.800pt}{2.200pt}}
\multiput(563.34,333.00)(-4.000,5.434){2}{\rule{0.800pt}{1.100pt}}
\put(558.34,343){\rule{0.800pt}{2.650pt}}
\multiput(559.34,343.00)(-2.000,5.500){2}{\rule{0.800pt}{1.325pt}}
\multiput(557.06,354.00)(-0.560,1.264){3}{\rule{0.135pt}{1.800pt}}
\multiput(557.34,354.00)(-5.000,6.264){2}{\rule{0.800pt}{0.900pt}}
\put(550.34,364){\rule{0.800pt}{2.400pt}}
\multiput(552.34,364.00)(-4.000,6.019){2}{\rule{0.800pt}{1.200pt}}
\multiput(548.07,375.00)(-0.536,0.909){5}{\rule{0.129pt}{1.533pt}}
\multiput(548.34,375.00)(-6.000,6.817){2}{\rule{0.800pt}{0.767pt}}
\multiput(542.07,385.00)(-0.536,1.020){5}{\rule{0.129pt}{1.667pt}}
\multiput(542.34,385.00)(-6.000,7.541){2}{\rule{0.800pt}{0.833pt}}
\multiput(536.08,396.00)(-0.520,0.627){9}{\rule{0.125pt}{1.200pt}}
\multiput(536.34,396.00)(-8.000,7.509){2}{\rule{0.800pt}{0.600pt}}
\multiput(525.52,407.40)(-0.543,0.514){13}{\rule{1.080pt}{0.124pt}}
\multiput(527.76,404.34)(-8.758,10.000){2}{\rule{0.540pt}{0.800pt}}
\multiput(513.74,417.39)(-0.685,0.536){5}{\rule{1.267pt}{0.129pt}}
\multiput(516.37,414.34)(-5.371,6.000){2}{\rule{0.633pt}{0.800pt}}
\multiput(504.19,423.38)(-1.096,0.560){3}{\rule{1.640pt}{0.135pt}}
\multiput(507.60,420.34)(-5.596,5.000){2}{\rule{0.820pt}{0.800pt}}
\multiput(493.86,428.38)(-1.432,0.560){3}{\rule{1.960pt}{0.135pt}}
\multiput(497.93,425.34)(-6.932,5.000){2}{\rule{0.980pt}{0.800pt}}
\put(482,431.84){\rule{2.168pt}{0.800pt}}
\multiput(486.50,430.34)(-4.500,3.000){2}{\rule{1.084pt}{0.800pt}}
\put(466,434.34){\rule{3.854pt}{0.800pt}}
\multiput(474.00,433.34)(-8.000,2.000){2}{\rule{1.927pt}{0.800pt}}
\put(574.0,301.0){\rule[-0.400pt]{0.800pt}{2.650pt}}
\put(434,434.34){\rule{3.854pt}{0.800pt}}
\multiput(442.00,435.34)(-8.000,-2.000){2}{\rule{1.927pt}{0.800pt}}
\put(425,431.84){\rule{2.168pt}{0.800pt}}
\multiput(429.50,433.34)(-4.500,-3.000){2}{\rule{1.084pt}{0.800pt}}
\multiput(415.54,430.06)(-1.768,-0.560){3}{\rule{2.280pt}{0.135pt}}
\multiput(420.27,430.34)(-8.268,-5.000){2}{\rule{1.140pt}{0.800pt}}
\multiput(404.53,425.06)(-1.264,-0.560){3}{\rule{1.800pt}{0.135pt}}
\multiput(408.26,425.34)(-6.264,-5.000){2}{\rule{0.900pt}{0.800pt}}
\multiput(396.19,420.07)(-0.797,-0.536){5}{\rule{1.400pt}{0.129pt}}
\multiput(399.09,420.34)(-6.094,-6.000){2}{\rule{0.700pt}{0.800pt}}
\multiput(387.85,414.08)(-0.654,-0.514){13}{\rule{1.240pt}{0.124pt}}
\multiput(390.43,414.34)(-10.426,-10.000){2}{\rule{0.620pt}{0.800pt}}
\multiput(375.85,404.08)(-0.487,-0.514){13}{\rule{1.000pt}{0.124pt}}
\multiput(377.92,404.34)(-7.924,-10.000){2}{\rule{0.500pt}{0.800pt}}
\multiput(368.08,391.11)(-0.516,-0.611){11}{\rule{0.124pt}{1.178pt}}
\multiput(368.34,393.56)(-9.000,-8.555){2}{\rule{0.800pt}{0.589pt}}
\multiput(359.08,380.02)(-0.520,-0.627){9}{\rule{0.125pt}{1.200pt}}
\multiput(359.34,382.51)(-8.000,-7.509){2}{\rule{0.800pt}{0.600pt}}
\multiput(351.07,368.08)(-0.536,-1.020){5}{\rule{0.129pt}{1.667pt}}
\multiput(351.34,371.54)(-6.000,-7.541){2}{\rule{0.800pt}{0.833pt}}
\multiput(345.07,357.63)(-0.536,-0.909){5}{\rule{0.129pt}{1.533pt}}
\multiput(345.34,360.82)(-6.000,-6.817){2}{\rule{0.800pt}{0.767pt}}
\put(337.34,343){\rule{0.800pt}{2.400pt}}
\multiput(339.34,349.02)(-4.000,-6.019){2}{\rule{0.800pt}{1.200pt}}
\put(333.34,333){\rule{0.800pt}{2.200pt}}
\multiput(335.34,338.43)(-4.000,-5.434){2}{\rule{0.800pt}{1.100pt}}
\multiput(331.07,326.08)(-0.536,-1.020){5}{\rule{0.129pt}{1.667pt}}
\multiput(331.34,329.54)(-6.000,-7.541){2}{\rule{0.800pt}{0.833pt}}
\put(323.84,312){\rule{0.800pt}{2.409pt}}
\multiput(325.34,317.00)(-3.000,-5.000){2}{\rule{0.800pt}{1.204pt}}
\put(450.0,437.0){\rule[-0.400pt]{3.854pt}{0.800pt}}
\put(321.84,291){\rule{0.800pt}{2.409pt}}
\multiput(322.34,296.00)(-1.000,-5.000){2}{\rule{0.800pt}{1.204pt}}
\put(320.34,281){\rule{0.800pt}{2.409pt}}
\multiput(321.34,286.00)(-2.000,-5.000){2}{\rule{0.800pt}{1.204pt}}
\put(324.0,301.0){\rule[-0.400pt]{0.800pt}{2.650pt}}
\put(319.84,260){\rule{0.800pt}{2.409pt}}
\multiput(319.34,265.00)(1.000,-5.000){2}{\rule{0.800pt}{1.204pt}}
\put(321.0,270.0){\rule[-0.400pt]{0.800pt}{2.650pt}}
\put(320.84,236){\rule{0.800pt}{3.132pt}}
\multiput(320.34,242.50)(1.000,-6.500){2}{\rule{0.800pt}{1.566pt}}
\put(323.34,218){\rule{0.800pt}{3.800pt}}
\multiput(321.34,228.11)(4.000,-10.113){2}{\rule{0.800pt}{1.900pt}}
\put(327.34,202){\rule{0.800pt}{3.400pt}}
\multiput(325.34,210.94)(4.000,-8.943){2}{\rule{0.800pt}{1.700pt}}
\put(331.34,192){\rule{0.800pt}{2.200pt}}
\multiput(329.34,197.43)(4.000,-5.434){2}{\rule{0.800pt}{1.100pt}}
\put(335.34,181){\rule{0.800pt}{2.400pt}}
\multiput(333.34,187.02)(4.000,-6.019){2}{\rule{0.800pt}{1.200pt}}
\multiput(340.39,174.63)(0.536,-0.909){5}{\rule{0.129pt}{1.533pt}}
\multiput(337.34,177.82)(6.000,-6.817){2}{\rule{0.800pt}{0.767pt}}
\multiput(346.39,164.63)(0.536,-0.909){5}{\rule{0.129pt}{1.533pt}}
\multiput(343.34,167.82)(6.000,-6.817){2}{\rule{0.800pt}{0.767pt}}
\multiput(352.40,156.52)(0.514,-0.543){13}{\rule{0.124pt}{1.080pt}}
\multiput(349.34,158.76)(10.000,-8.758){2}{\rule{0.800pt}{0.540pt}}
\multiput(361.00,148.08)(0.639,-0.512){15}{\rule{1.218pt}{0.123pt}}
\multiput(361.00,148.34)(11.472,-11.000){2}{\rule{0.609pt}{0.800pt}}
\multiput(375.00,137.06)(1.264,-0.560){3}{\rule{1.800pt}{0.135pt}}
\multiput(375.00,137.34)(6.264,-5.000){2}{\rule{0.900pt}{0.800pt}}
\put(385,131.34){\rule{1.686pt}{0.800pt}}
\multiput(385.00,132.34)(3.500,-2.000){2}{\rule{0.843pt}{0.800pt}}
\put(392,129.34){\rule{1.686pt}{0.800pt}}
\multiput(392.00,130.34)(3.500,-2.000){2}{\rule{0.843pt}{0.800pt}}
\put(399,127.84){\rule{2.168pt}{0.800pt}}
\multiput(399.00,128.34)(4.500,-1.000){2}{\rule{1.084pt}{0.800pt}}
\put(322.0,249.0){\rule[-0.400pt]{0.800pt}{2.650pt}}
\put(408.0,129.0){\rule[-0.400pt]{2.409pt}{0.800pt}}
\put(223,448){\usebox{\plotpoint}}
\multiput(223.00,446.09)(0.984,-0.500){411}{\rule{1.773pt}{0.121pt}}
\multiput(223.00,446.34)(407.320,-209.000){2}{\rule{0.887pt}{0.800pt}}
\sbox{\plotpoint}{\rule[-0.200pt]{0.400pt}{0.400pt}}%
\put(438,307){\makebox(0,0){$+$}}
\end{picture}

%% file: rasgf.bbl
\begin{thebibliography}{99}
\bibitem{bah} Bahcall N.A., Fan X., Cen R., 1997, ApJ 485, L53
\bibitem{bon} Bond J.R., Jaffe A.H., Phil. Trans. R. Soc. Lond. A 
 (in press) astro-ph/9809043
\bibitem{bor} Borgani S. et al., 1999, astro-ph/9908155
\bibitem{bro} Broadhurst T., Jaffe A.H., astro-ph/9904348
\bibitem{cha} Chaboyer B., 1998, (subm. to Elsevier) astro-ph/9808200
\bibitem{chi} Chiba M., Yoshii Y., 1999, ApJ 510, 42
\bibitem{dal} Daly R.A., Guerra E.J., Wan Lin, astro-ph/9803265
\bibitem{don} Donahue M., Voit G.M., 1999, ApJ 523, L137
\bibitem{eke} Eke V.R. et al., 1998, MNRAS 298, 1145
\bibitem{evr} Evrard A.E., 1997, MNRAS 292, 289
\bibitem{fal} Falco E.E., Kochanek C.S., Munoz J.M., 1998, ApJ 494, 47
\bibitem{imm} Im M., Griffiths R.E., Ratnatunga K.U., 1997, ApJ 475, 457
\bibitem{jam} James F., Roos M., 1975, Comput. Phys. Comm. 10, 343
\bibitem{jim} Jimenez R., (IOP in press) astro-ph/9810311
\bibitem{lin} Lineweaver Ch.H., 1998, ApJ 505, L69
\bibitem{mar} Markevitch, M., 1998, ApJ 504, 27
\bibitem{mel} Melchiorri A. et al., astro-ph/9911445
\bibitem{mou} Mould J. R. et al., astro-ph/9909260
\bibitem{nev} Nevalainen J., Roos M., 1998, A\&A 339, 7
\bibitem{per} Perlmutter S et al., 1998, Nature 391, 51; 1999, ApJ 517, 565
\bibitem{rie} Riess A.G. et al., 1998, AJ 116, 1009
\bibitem{ro1} Roos M., Harun-or-Rashid S.M., 1998, A\&A 329, L17
\bibitem{ro2} Roos M., Harun-or-Rashid S.M., 1999, astro-ph/9901234
\bibitem{rou} Roukema B.F., Mamon G.A., astro-ph/9911413
\bibitem{teg} Tegmark M., 1999, ApJ 514, L69 
\bibitem{tot} Totani T., Yoshii Y., Sato K., 1997, ApJ 483, L75
\bibitem{tyt} Tytler D., Fan X.-M., Burles S., 1996, Nature 381 207
\bibitem{wei} Weinberg D.H. et al., 1999 ApJ 522, 563
\bibitem{wil} Willick J.A. et al., 1997, ApJ 486, 629; 1999, Astrophys. J. S. 109, 333
\bibitem{zeh} Zehavi I., Dekel A., 1999, preprint astro-ph/9904221

\end{thebibliography}
